\documentstyle[aps,twocolumn,epsf]{revtex}

\def\nle{\ \raise.3ex\hbox{$<$}\kern-0.8em\lower.7ex\hbox{$\sim$}\ }
\def\nge{\ \raise.3ex\hbox{$>$}\kern-0.8em\lower.7ex\hbox{$\sim$}\ }
\def\chihis{\chi_{\rm ZFC}(L(\tau),R(t_{\rm w}))}
\def\femntio{Fe$_{0.50}$Mn$_{0.50}$TiO$_3$}
\def\mFCT{m_{\rm FC}(T)}
\def\mZFCT{m_{\rm ZFC}(T)}

\def\mgt{m_{\rm ZFC}(t)}
\def\mgttwT{m_{{\rm ZFC},T}(t_a,t_{\rm w})}
\def\mgTtau{m_{{\rm ZFC},T[\tau]}(t_a,t_{\rm w})}
\def\Tc{T_{\rm c}}
\def\tw{t_{\rm w}}
\def\taueffi{\tau_i^{\rm eff}}

\begin{document}
\draft
\twocolumn[
\hsize\textwidth\columnwidth\hsize\csname@twocolumnfalse\endcsname

\title{Aging of the Zero-Field-Cooled Magnetization in Ising Spin
 Glasses:\\ 
Experiment and Numerical Simulation}

\author{Lorenzo W. Bernardi$^{1}$, Hajime Yoshino$^{1}$, Koji Hukushima$^{1}$,
Hajime Takayama$^{1}$, Aya Tobo$^{2}$, and Atsuko Ito$^{3}$}

\address{$^{1}$ Institute for Solid State Physics , University of Tokyo, 
  Kashiwa, 277-8581, Japan\\
$^{2}$Institute of Materials Research, Tohoku University, Sendai 
980-8577, Japan\\
$^{3}$The institute of Physical and Chemical Research (RIKEN), Wako, Saitama,
351-0198, Japan}
\date{May 30, 2000}
\maketitle
\begin{abstract}
A new protocol of the zero-field-cooled (ZFC) magnetization
process is studied experimentally on an Ising spin-glass (SG) \femntio\ 
and numerically on the Edwards-Anderson Ising SG model.
Although the time scales differ very much between the experiment and the 
simulation, the behavior of the ZFC magnetization observed in the two
systems can be interpreted by means of a common scaling expression based
on the droplet picture. The results strongly suggest that the SG
coherence length, or the mean size of droplet excitations, involved even 
in the experimental ZFC process, is about a hundred lattice distances or
less. 
\end{abstract}
\vskip.5pc ] 
\narrowtext

\sloppy

Spin glasses generally exhibit puzzling aging behavior such that their 
magnetic response in the spin-glass (SG) phase below its transition 
temperature $\Tc$ depends crucially on their thermal history within 
the phase. For example, an isothermal zero-field-cooled (ZFC) magnetization, 
$\mgttwT$, is defined as the magnetization induced by a magnetic field 
$h$ which is applied {\it after} a system is rapidly quenched from 
above $\Tc$ to a temperature $T$ below $\Tc$ and is aged for a waiting 
time of $\tw$, and $t_a=\tau+\tw$ with $\tau$ being the time that the 
system ages at $T$ under $h$. With increasing $\tau$, $\mgttwT$ grows
slowly. In contrast to ideal equilibrium situations, however, it depends
not only on $\tau$ but also on the waiting time $\tw$, namely, the
larger $\tw$, the slower the growth of $\mgttwT$. This was the first
observation of aging phenomena in spin glasses \cite{Lundgren}. Since
then most of the studies have focused on its $\tw$-dependence.

In the present work, on the other hand, we have studied the ZFC 
magnetization focusing on its dependence on thermal processes after 
the field $h$ is applied.  We denote it by $\mgTtau$ where $\tw$ 
symbolically represents a common thermal history before $h$ 
is applied, while $T[\tau]$ does various heating processes after $h$ 
is switched on, e.g., with a fixed heating rate but with an 
intermittent stop(s) at a different temperature(s). We have carried 
out this new protocol of the ZFC magnetization measurement both 
experimentally on an Ising SG material \femntio \cite{Ito}, 
hereafter referred to as FeMnTiO, and numerically on the 3D Gaussian
Edwards-Anderson (EA) Ising SG model. Quite interestingly, we have found
that $\mgTtau$ measured in the two systems are well interpreted by a
unified way, or more explicitly, they are described by a common scaling
expression which can be derived based on the droplet picture for the SG 
phase \cite{FH}. The purpose of the present letter is to demonstrate 
this finding and to discuss its implications on the aging dynamics in
spin glasses.

According to the droplet picture, the mean size, $R(\tw)$, of 
domains of an equilibrium state grows continuously when a spin 
glass relaxes with increasing $\tw$. In fact its growth law in 
isothermal aging at a temperature $T$, denoted by $R_T(\tw)$, 
has been numerically studied by the various 
groups \cite{Huse,Kisker,Marinari-98,KYT1}, and within the numerical 
accuracy all the results are consistent with each other. Here we 
quote our result\cite {KYT1},
\begin{equation}
R_T(\tw) \simeq (\tw/\tau_0)^{b_{\rm f}T/\Tc}, 
\label{eq:Rt}
\end{equation}
where $\tau_0$ is the microscopic unit of time which is one Monte 
Carlo step (MCS) in the simulation, and $b_{\rm f} \simeq 
0.17\Tc/J$ \cite{KYT1} $\simeq 0.153$ with $\Tc \simeq
0.9J$ \cite{EATc}, $J$ being the variance of the interactions. 
Furthermore we have shown~\cite{KYT2} that the aging properties of the 
spin auto-correlation function are consistently explained when the 
same functional form is adopted for the maximum size of droplets
which can be activated within a time scale of $\tau$, $L_T(\tau)$, i.e.,
\begin{equation}
L_T(\tau) \simeq (\tau/\tau_0)^{b_{\rm f}T/\Tc} 
\label{eq:Ltau}
\end{equation}
with the same constant $b_{\rm f}$. For an aging process in which 
the temperature is changed as in our new protocol, the maximum 
size of droplet excitations at time $t_a=\tau+\tw$, denoted 
by $L(\tau)$, is evaluated by the successive use of eq.(\ref{eq:Ltau}) 
as described by eqs.(\ref{eq:scale1}) and (\ref{eq:scale2}) below. 

In terms of $L(\tau)$ introduced just above the common scaling
expression we have found is written as 
\begin{equation}
h^{-1} \mgTtau = \chihis,
\label{eq:eq1}
\end{equation}
where $R(\tw)$ is the mean domain size at the instance when the field
$h$ is switched on. 
For isothermal processes under a field within the linear response 
regime the droplet theory~\cite{FH} has already predicted a similar 
scaling form, i.e., $\mgttwT$/h is a function of $R_T(\tw)$ and 
$L_T(\tau)$. Equation (\ref{eq:eq1}) extends it to aging processes 
which involve temperature changes. Also, by the low temperature 
expansion of the droplet theory, $\mgttwT$/h in the 
quasi-equilibrium regime with $R_T(\tw) \gg L_T(\tau)$ is 
shown~\cite{FH,KYT2} not to depend explicitly on $T$. Equation 
(\ref{eq:eq1}) shares the same property but for the whole range of 
$\tau$. In relation to this it is worth pointing out that the 
field-cooled (FC) magnetization, $\mFCT$, which is supposed to be 
lim$_{t_a \rightarrow \infty}\mgttwT$, is nearly independent of 
$T$ for the Ising spin glasses such as the FeMnTiO and the EA Ising
model. 

The unified description of the ZFC magnetizations means that 
eqs.(\ref{eq:eq1}) and (\ref{eq:Ltau}) are quantitatively consistent
with the measured results on the FeMnTiO spin glass when $\tau_0 \sim
10^{-12}$s and the same value of $b_{\rm f}$ as that of the EA-SG
model are used. It is, to our knowledge, the first case in the study on 
aging in spin glasses that such a {\it quantitative} agreement is
obtained in the direct comparison of the experiment with the numerical 
simulation. 

\begin{figure}
\leavevmode\epsfxsize=70mm
\epsfbox{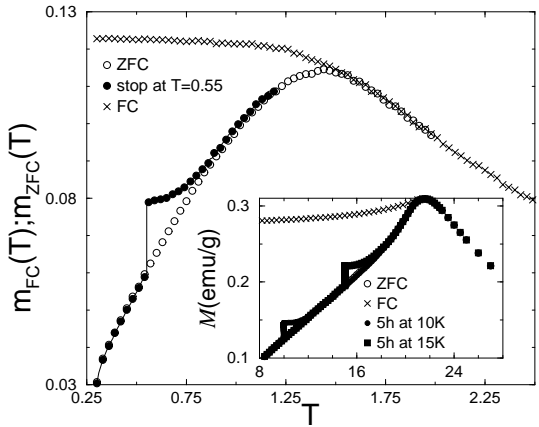}
\caption{The ZFC ($\circ$) and FC ($\times$) magnetizations of 
the EA Ising SG model. The solid circles represent $\mZFCT$ with an
 intermittent stop of $10^4$ MCS at $T=0.55$. 
In the inset we show the 
 corresponding results measured in the Ising SG \femntio. 
 The solid circles and squares represent  $\mZFCT$ with an 
intermittent stop of 5 hours at 10.0K and 15.0K, respectively. } 
\label{fig:Fig1}
\end{figure}

We have measured the induced magnetization of the single crystal 
FeMnTiO sample with $\Tc=$21.5K \cite{Itoref5} by applying the field 
parallel to the hexagonal $c$-axis of the sample (see \cite{Ito} for 
further experimental details). The numerical simulations have been 
done on the 3D EA model with nearest neighbour interactions of mean 
zero and variance $J$ which is the unit of energy. We have used the 
heat-bath dynamics and have simulated several hundred samples with a 
linear size $L=24$ to make 
the error bars less than the size of the symbol in each plot. It is 
noted that the units of time and temperature used in the present work 
are respectively second and Kelvin in the experiment, and MCS and $J$, 
which we put unity below, in the simulation. Our processes measuring 
magnetizations in the experiment [simulation] are as follows. The 
system was first cooled from 50K to 5K [2.5$J$ to 0.3$J$] at a rate of 
1.0 K/min [0.05$J$/10MCS] under zero field. Once the lowest 
temperature was reached the field of 100 Oe [0.2$J$] was applied and 
the sample was heated stepwise with a step of height 0.2K [0.01$J$] 
and length 100s [100MCS]. The last 13s [10MCS] of the latter were used 
to measure $\mgTtau$ which we regard as the ZFC magnetization at that 
temperature, $\mZFCT$. At 50K, in the experiment, we turned to 
decrease the temperature and measured the FC magnetization, $\mFCT$, on 
cooling with the same rate and in the same field. In the simulation 
$\mFCT$ was measured on cooling the sample under $h=0.2J$ from 2.5$J$
down to 0.3$J$ with the same rate as for $\mZFCT$. 

The ZFC and FC magnetizations as well as the one with an intermittent 
stop of the real and model Ising spin glasses, shown in  
Fig.~\ref{fig:Fig1}, look qualitatively quite similar to each other. 
Particularly, $\mFCT$ of both systems become nearly flat at $T$ 
sufficiently lower than $\Tc$. We can also observe the following common 
feature in $\mZFCT$ with an intermittent stop. During the stop $\mZFCT$ 
increases, and when the heating is restarted, it continues to increase. 
However it increases significantly much slowly than $\mZFCT$ without 
the stop, and is caught up by the latter at a certain $T$ below $\Tc$. 

\begin{figure}
\leavevmode\epsfxsize=70mm
\epsfbox{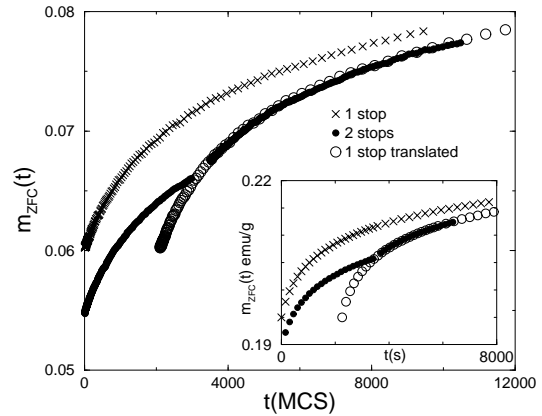}
\caption{Comparison of $m_{{\rm ZFC},T[\tau]}(t_a,t_{\rm w})$ with one
 and two intermittent stops in the EA model (main frame) and the FeMnTiO 
 (inset) [2].  Here they are denoted as $m_{\rm ZFC}(t)$ (see the 
 text).}
\label{fig:Fig2}
\end{figure}

In the inset [main frame] of Fig.~\ref{fig:Fig2} we demonstrate the most
interesting observation in our new protocol. It is the comparison of
the explicit time evolutions of $\mgTtau$ of the FeMnTiO spin glass [the 
EA SG model] with one intermittent stop at $T=15.0$K [0.55$J$] and two
stops at $T=14.6$K by 1 hour and at $T=15.0$K [$T=0.50J$ by 3000 MCS and
at $T=0.55J$]. Here we rewrite them as $\mgt$ since we are interested in
their time evolution. The origin of the time $t$ is set at the
instance when the system is heated up to 15.0K [0.55$J$] for the 1-stop
process and to 14.6K [0.50$J$] for the 2-stop process. It is seen that,
when the 1-stop curve (crosses) is shifted to the right by 2250 s [2100
MCS] (open circles), it almost perfectly collapses to the branch of
$T=15.0$K [0.55$J$] of the 2-stop curve (solid circles). In other words, 
once $\mgTtau$ of two different thermal processes coincide with each
other, their subsequent evolutions under a common condition of $T$ and
$h$ are also identical. Keeping in mind that these processes have a
common $R(\tw)$, or a common thermal history before $h$ is applied, the 
results on both the real and model spin glasses are consistent with
eq.(\ref{eq:eq1}) derived from the droplet picture. 

\begin{figure}
\leavevmode\epsfxsize=70mm
\epsfbox{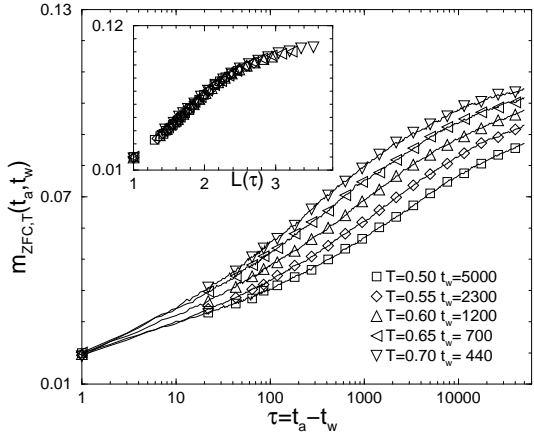}
\caption{Evolution of the isothermal ZFC magnetization of the EA model
 at different temperatures but starting from a common 
 $R_T(\tw)\ (\simeq 2.1)$. In the inset, the same data are plotted
 against $L_T(\tau)$ evaluated 
 by eq.(2).}
\label{fig:Fig3}
\end{figure}

To ascertain the scaling expression of eq.(\ref{eq:eq1}), and to
demonstrate that the $T$-dependence of the scaling function $\chihis$
comes out only through that of $L(\tau)$, we have performed the 
simulation of isothermal ZFC magnetization processes at various
temperatures but by choosing the waiting time $\tw$ after the
instantaneous quench such that $R_T(\tw)$ given by eq.(\ref{eq:Rt})
become the same for all temperatures. The results are shown in 
Fig.~\ref{fig:Fig3} where $\mgttwT$ are plotted against the time 
$\tau$ after the field $h=0.2$ is switched on. The chosen $R_T(\tw)$ is 
$\simeq 2.1$ and the corresponding $\tw$ at each temperature are
indicated in the figure. 

The evolutions of $\mgttwT$ in Fig.~\ref{fig:Fig3} depend on $T$ except 
at $\tau \simeq 1$. When, however, the same data are plotted against
$L_T(\tau)$ evaluated by eq.(\ref{eq:Ltau}), they are nicely superposed 
upon each other as shown in the inset of Fig.~\ref{fig:Fig3}. This result 
strongly supports the scaling expression of eq.(\ref{eq:eq1}) in the 
following two respects: the thermal history {\it before} the field is 
applied is in fact renormalized to $R(\tw)$, and the size $L_T(\tau)$ of 
droplet excitations which are dominant in the time scale $\tau$ {\it
after} the field is applied is properly given by eq.(\ref{eq:Ltau}). We
also note that, within the accuracy of the present numerical analysis,
the scaling behavior is ascertained not only in the quasi-equilibrium
regime, where $L_T(\tau)\nle R_T(\tw)$ or $\tau \ll \tw$, as predicted
by the droplet theory~\cite{FH,KYT2}, but also in the
out-of-equilibrium, or aging regime, where $L_T(\tau)\nge R_T(\tw)$ or
$\tau \gg \tw$. In the latter regime $L_T(\tau)$ plays a role of the
mean domain size at time $t_a$ in place of $R_T(\tw)$.  

Lastly let us extend the scaling analysis done above to our new
experimental protocol in which the temperature is changed. For this
purpose it is important to note that, so long as the system is
in the SG phase, the mean domain size $R(\tw)$ of the EA model analyzed
numerically ever grows continuously even when the temperature is changed 
discontinuously \cite{KYT4}. It is then natural for us to expect that
this is also the case for $L(\tau)$ under a fixed field $h$, so that we
may successively use eq.(\ref{eq:Ltau}) as follows. Suppose that
temperature is changed as $T_i=T_{i-1}+\Delta T_i$ and a period of time
$d\tau_i$ is spent at each temperature $T_i$. We assume that the length
scale achieved after the step at $T_i$ is given as
\begin{equation}
L_i = (\taueffi +d\tau_i)^{b_{\rm f}T_i/\Tc},
\label{eq:scale1}
\end{equation}
where $\taueffi$ is an effective elapsed time to take 
into account the thermal history up to the instance just before the
temperature is changed to $T_i$. It is the time needed to achieve the
length scale $L_{i-1}$ reached after the $(i-1)$-th step by isothermal 
aging at $T_i$,
\begin{equation}
(\taueffi)^{b_{\rm f}T_i/\Tc}=L_{i-1}.
\label{eq:scale2}
\end{equation}
Combining these we obtain a recursion formula for $L_i$ and/or
$\taueffi$ which is solved for any given temperature process with
$L(\tau=0)=0$. 

\begin{figure}
\leavevmode\epsfxsize=70mm
\epsfbox{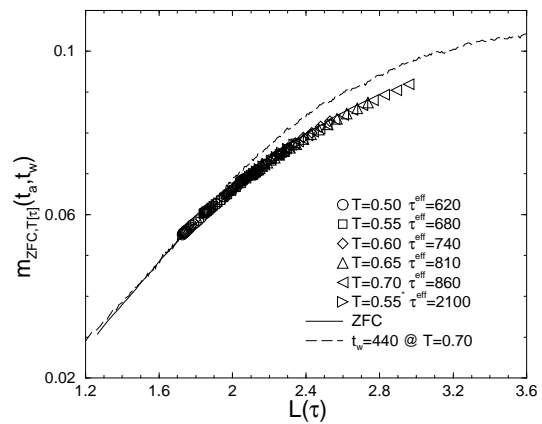}
\caption{Scaling plot of $\mgTtau$ of the EA model against $L(\tau)$
 evaluated by eqs.(\ref{eq:scale1}) and (\ref{eq:scale2}). The solid
 line represents the one without intermittent stop, while the symbols
 are those at the stop at various temperatures $T$ where $\tau^{\rm eff}$ 
is the effective elapsed time at the instance when the system just
 reaches $T$.  The data at $T=0.55$  with $\tau^{\rm eff}=2100$ MCS
 represent the second stop in the 2-stop  process. The dashed line is
 one of the curves shown in the inset of
 Fig. 3.}
\label{fig:Fig4}
\end{figure}

We have first applied the above-mentioned analysis to $\mgTtau$ of the
EA model in the processes with $T$-changes shown in Figs.~\ref{fig:Fig1}
and \ref{fig:Fig2}. The results are shown in Fig.~\ref{fig:Fig4}. All
$\mgTtau$ nicely lie on a universal curve, indicating that the scaling
of eq.(\ref{eq:eq1}), combined with the above analysis on $L(\tau)$,
does work well. In the figure we also draw one of the curves in the
isothermal process shown in the inset of Fig.~\ref{fig:Fig3}. It has the 
different thermal history before the field is applied, i.e., the
different $R(\tw)$, from those of the $T$-change process. It
significantly deviates from the universal curve for the latter as
expected from eq.(\ref{eq:eq1}).

\begin{figure}
\leavevmode\epsfxsize=80mm
\epsfbox{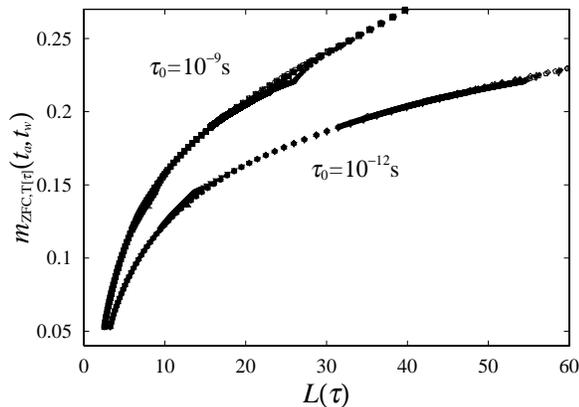}
\caption{Scaling plots of $\mgTtau$ of the FeMnTiO spin glass shown in 
the inset of Figs. 1 and 2. The abscissa 
$L(\tau)$ is evaluated by eq.(2) with $\tau_0=10^{-9}$ s 
and $10^{-12}$ s.}
\label{fig:Fig5}
\end{figure}

Surprisingly, the $\mgTtau$ curves measured in the FeMnTiO spin glass and 
shown in the inset of Figs.~\ref{fig:Fig1} and \ref{fig:Fig2} obey the 
same scaling law as those of the EA model. Indeed, their scaling plots 
demonstrated in Fig.~\ref{fig:Fig5} are obtained by making use of 
eq.(\ref{eq:Ltau}) with the identical exponent of $b_{\rm f}T/\Tc$ but
with $\tau_0=10^{-9}$ s and $10^{-12}$ s, instead of $\tau_0=1$ MCS for
the EA model. The two portions with the dense data points at around 
$\mgTtau \simeq 0.14$ and $0.20$ represent the results with intermittent 
stops at 9.6K, 10.0K and at 14.6K, 15.0K, respectively. At a glance we
see that the scaling with $\tau_0=10^{-12}$ s is the better. We have not 
performed further analysis on the sensitivity of the collapse of the 
curves to the parameter values, because the evolution law of 
eq.(\ref{eq:Ltau}) or (\ref{eq:Rt}) already involves certain 
uncertainty. For example, $R_T(\tw)$ extracted by the ac
susceptibility in a sufficiently wide frequency range is better
expressed by a function of ln$\tw$ \cite{Saclay-90}. Furthermore we are
faced to similar but more serious problems in the temperature range
close to $\Tc$, such as the crossover to the critical regime 
\cite{FH,HYT}, whose analysis has been omitted in the present work.
However we consider that the scaling results obtained above are enough
for us to claim that the droplet picture expressed by eq.(\ref{eq:eq1})
is consistent with the aging behavior of the ZFC magnetization in the
FeMnTiO Ising spin glass, and that the length scale involved in the
process is about a hundred lattice distances or less. They also imply
the continuous growth of $L(\tau)$ and so $R(\tw)$, without suffering
from the chaotic effect \cite{BMchaos}, at least in the present 
experimental protocol.

Another interesting result of the present work is that the evolution 
law of $L_T(\tau)$ given by eq.(\ref{eq:Ltau}), or the growth law of 
$R_T(\tw)$ given by eq.(\ref{eq:Rt}), almost coincides with the growth 
law of the SG coherence length,  $\xi_T(t)$, which Joh 
{\it et al.}~\cite{Joh1,Joh2} has extracted from the thermoremanent 
magnetization (TRM) of the spin glasses CuMn, AgMn and 
CdIn$_{0.3}$Cr$_{1.7}$S$_4$, though none of them is an Ising spin
glass. Their results imply that 
the evolution law of $L_T(\tau)$ depends on 
the field at least in the the crossover region between the 
quasi-equilibrium and the aging regimes. In the present work, however, 
such a field effect has been neglected regarding it as a higher order 
effect. 

To conclude we have demonstrated that the ZFC magnetizations measured
in both real and model Ising spin glasses are described by a common 
scaling expression which is consistent with the droplet picture on the 
SG aging dynamics. The result indicates that the numerical simulation 
on the model spin glass supplies us not only qualitative but also 
quantitative information on aging phenomena in real spin glasses.

This work is supported by Grant-in-Aids for International Scientific
Research Program (\#10044064) and for Scientific Research Program 
(\#10640362), from the Ministry of Education, Science, Sports and
Culture, Japan. The present simulation has been performed by the
computing facilities at Supercomputer Center, Institute for Solid State 
Physics, University of Tokyo.

\end{document}